\documentclass[prb,twocolumn,amsmath,showpacs]{revtex4}

\usepackage{graphicx}% Include figure files

\begin{document}
\input epsf

\title {Reinterpretation of the equilibrium magnetization 
of a Tl$_{2}$Ba$_{2}$CaCu$_{2}$O$_{8+x}$ single crystal. Another phase 
transition in the mixed state of high-$T_{c}$ superconductors?}

\author {I. L. Landau$^{1,2}$ H. R. Ott$^{1}$}
\affiliation{$^{1}$Laboratorium f\"ur Festk\"orperphysik, ETH H\"onggerberg, 
CH-8093 Z\"urich, Switzerland}
\affiliation{$^{2}$Kapitza Institute for Physical Problems, 117334 Moscow, 
Russia}

\date{\today}

\begin{abstract}
We apply a recently developed scaling procedure for the analysis of the 
equilibrium magnetization $M$ that was measured on a 
Tl$_{2}$Ba$_{2}$CaCu$_{2}$O$_{8+x}$ single crystal and was recently 
reported in the literature. The results of our analysis are distinctly 
different from those obtained in the original publication where the 
Hao-Clem model served to analyze the magnetization data. We argue  
that the Hao-Clem model is not adequate for a quantitative description 
of the mixed state in high-$T_{c}$ superconductors especially in high 
magnetic fields. The scaled equilibrium magnetization data reveal a 
pronounced kink in the $M(H)$ dependence that might be indicative 
of a phase transition in the mixed state.  
\end{abstract}

\maketitle

First, we briefly review the current situation in the interpretation of 
magnetization measurements in the mixed state of high-$T_{c}$ (HTSC) 
superconductors. The most widely used approach for evaluating the 
thermodynamic critical field $H_{c}$ and the Ginzburg-Landau (GL) 
parameter $\kappa$ of HTSC's from magnetization data is the Hao-Clem 
model. \cite{2,3} Using the values of $H_{c}$ and $\kappa$ that result 
from the field dependence of the magnetization measured at different 
temperatures, the temperature dependence of the upper critical field 
$H_{c2}(T)$, as well as the magnetic field penetration depth 
$\lambda (T)$, and the superconducting coherence length $\xi (T)$ may be 
calculated. However a recent theoretical analysis of the Hao-Clem 
approach demonstrated that the calculation of the local magnetic 
induction by a linear superposition of contributions of individual 
vortices, as it was done in Ref. \onlinecite{3}, is quantitatively 
acceptable only for low magnetic fields. \cite{pog1, pog2} A rather 
direct confirmation of the inadequecy of the Hao-Clem model for the 
quantitative analysis of experimental results is provided by Figs. 3(a) 
and 3(b) in Ref. \onlinecite{4} 
where data on the equilibrium magnetization of Tl-based single 
crystals were presented and analyzed. The quoted Fig. 3(a) reveals a 
rather strong increase of $\kappa$ with increasing temperature, 
resulting in a very unusual temperature dependence of the upper critical 
field with $H_{c2}$ being practically temperature independent at low 
temperatures and almost diverging at temperatures above 85-90 K (Fig. 
3(b)). Because this anomalous increase of $H_{c2}$ starts well below the 
superconducting critical temperature $T_{c}$ of the investigated sample, 
it can hardly be ascribed to fluctuation effects. This very strong 
increase of $\kappa$ upon approaching the critical temperature is not a 
feature that was observed in one particular study. It turns out to be a 
rather universal result if the Hao-Clem model is employed for the 
analysis of magnetization data. \cite{hc1, hc2, hc3, hc4, hc5, hc6, hc7, 
hc8, hc9, hc10, hc11, hc12} We argue that the magnetic field dependence 
of the magnetization at constant temperature provided by the Hao-Clem 
model is not sufficiently accurate and therefore, a fitting of the 
relevant theoretical formula to experimental data results in an 
unphysical temperature dependence of the GL parameter $\kappa$, one 
of the key parameters in the Hao-Clem model. In the GL theory $\kappa$ is 
temperature independent and it only slowly decreases with increasing 
temperature in the framework of microscopic theories. \cite{th1, 
th2} Below we show that the quoted experimental data may very well be 
interpreted by assuming that $\kappa$ is temperature independent. The 
temperature dependence of $H_{c2}$ which results from our analysis, is 
in good agreement with expectations of the GL theory. 

Our analysis of the original data which were presented in Ref. 
\onlinecite{4} is based on a scaling procedure developed in Ref. 
\onlinecite{5}. It was shown that if the GL parameter $\kappa$ is 
temperature independent, the equilibrium magnetizations in the mixed 
state of type-II superconductors at two different temperatures $T_{0}$ 
and $T$ are related by 
%%%%
\begin{equation}
M(H/h_{c2},T_0)=M(H,T)/h_{c2}+c_0(T)H,
\end{equation}
%%%%
with $h_{c2}(T) = H_{c2}(T)/H_{c2}(T_{0})$ representing a normalized 
upper critical field. \cite{5} The first term on the right-hand side of 
Eq. (1) is universal for any type-II superconductor, while the second one 
is introduced to account for the temperature dependent paramagnetism of 
HTSCÕs in the normal state. \cite{6} In the following we use $M_{eff}$ 
to denote the magnetization calculated using Eq. (1) in order to 
distinguish it from direct experimentally measured values of $M$. The 
quantities $h_{c2}(T)$ and $c_{0}(T)$ are the adjustable parameters of 
the scaling procedure and they are established from the condition that 
the $M_{eff}$ curves, calculated from the $M(H)$ measurements in the 
reversible regime at different temperatures, collapse onto a single curve, 
representing the equilibrium magnetization at $T = T_{0}$ (see Ref. 
\onlinecite{5} for details). 

It has been shown that the scaling procedure described by Eq. (1) works 
rather well for different HTSC materials including single crystals and 
grain-aligned samples, \cite{5} as well as polycrystals and ceramics. 
\cite{8} An important advantage of this approach is that no particular 
formula for $M(H)$ has to be assumed {\it a priori} and therefore, this 
scaling procedure may be used for any type-II superconductor, independent 
of the pairing type, the absolute value of $\kappa$, the anisotropy of 
superconducting parameters, or the demagnetization factor of the sample.
%%%%%%%%%%
\begin{figure}[t]
 \begin{center}
  \epsfxsize=1\columnwidth \epsfbox {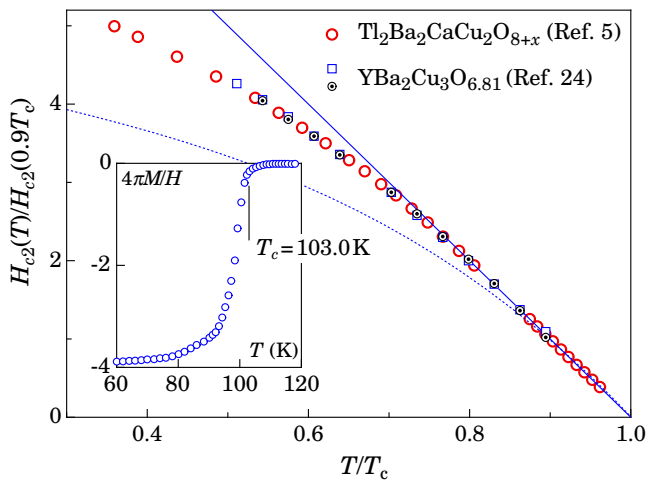}
  \caption{The dependence of the normalized upper critical field on 
           $T/T_{c}$. The solid line is the best linear fit to the data 
           points for $T \ge 90$ K. The dotted line represents the 
           \textquotedblleft universal\textquotedblright 
           $h_{c2}(T/T_{c})$ curve reported in Ref. \onlinecite{5} 
           for most of the HTSC compounds. The data points for two 
           underdoped YBCO samples with the same oxygen content, 
           investigated in Ref. \onlinecite{10}, were taken from Ref. 
           \onlinecite{8}. The inset shows the $M(T)$ curve measured at 
           $H = 10$ Oe. The short vertical line indicates the position 
           of $T_{c}$ as evaluated by the linear extrapolation of 
           $h_{c2}(T)$ to $h_{c2} = 0$.}
 \end{center}
\end{figure}
%%%%%%%%%%

As has briefly been outlined above (see Ref. \onlinecite{5} for details), 
Eq. (1) serves to evaluate the temperature variation of the normalized upper 
critical field $h_{c2}(T)$. It turns out that $h_{c2}(T)$ for this Tl2212 
sample varies perfectly linearly with $T$ above 90 K. This linearity allows 
for a quite accurate evaluation of the critical temperature $T_{c}$ by 
extrapolating the $h_{c2}(T)$ curve to $h_{c2} = 0$. The inset of Fig. 
1 demonstrates that the value of $T_{c}$ resulting from this procedure 
is consistent with the temperature dependence of the low-field 
magnetization of this sample. The normalized upper critical field 
$h_{c2}$ versus $T/T_{c}$ is shown in the mainframe of Fig. 1.

As has been shown in our previous studies,\cite{5,8} all investigated 
HTSC's may so far be divided into two groups. For each group, the 
$h_{c2}(T/T_{c})$ curves for the individual HTSC compounds turn out to 
be identical but the as yet identified two universal curves are distinctly 
different. The major group includes practically all different families 
of HTSCÕs, except Y$_{2}$Ba$_{4}$Cu$_{7}$O$_{15}$ and underdoped 
YBa$_{2}$Cu$_{3}$O$_{7-x}$ (YBCO) compounds which belong to the other group. 
As may be seen in Fig. 1, $h_{c2}(T)$ of this particular Tl2212 sample 
is practically identical with the $h_{c2}(T)$ curves of two underdoped YBCO 
samples investigated in Ref. \onlinecite{10}. This implies that the smaller 
group is not limited to underdoped Y-based cuprates. It was shown in Ref. 
\onlinecite{8} that a ceramic Tl2212 sample with $T_{c} = 107$ K, which 
was investigated in Ref. \onlinecite{11}, belongs to the other group 
represented by the dotted line in Fig. 1. A possible explanation for this 
difference between these two Tl-based samples may be sought in the 
difference of the oxygen content. 

%%%%%%%%%%
\begin{figure}[ht!]
 \begin{center}
  \epsfxsize=1\columnwidth \epsfbox {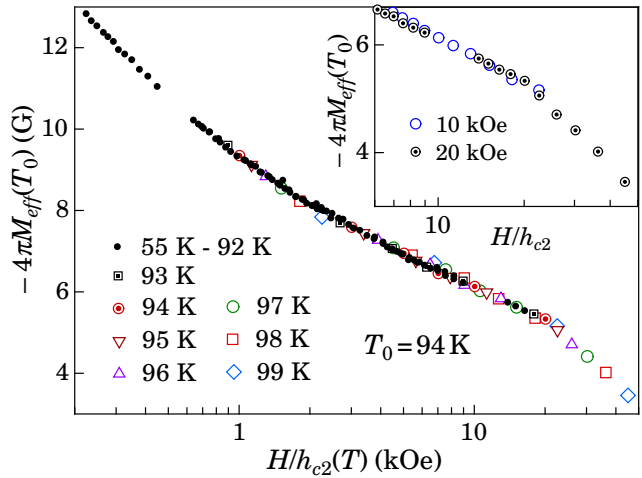}
  \caption{The scaled magnetization $M_{eff}$ calculated for $T_{0} = 94$ 
           K. The inset emphasizes the high field part of $M_{eff}$.} 
 \end{center}
\end{figure}
%%%%%%%%%%
The dependence of $M_{eff}(T_{0})$ on $H/h_{c2}(T)$ that results from our 
scaling procedure is shown in Fig. 2. Because of the high quality of the 
experimental data presented in Ref. \onlinecite{4} and the extended covered 
range of magnetic fields, the scaling is nearly perfect. The $M_{eff}(H)$ 
data points, calculated from the measurements at different temperatures, 
combine to a single curve with virtually no deviations. The remarkable 
feature of this curve is a pronounced kink at $H/h_{c2} \approx 20$ kOe. 
This kink clearly indicates a significant change in the properties of the 
mixed state. Unfortunately, the measurements in Ref. \onlinecite{4} are 
limited to magnetic fields $H \le 20$ kOe and, as may be seen in Fig. 2, 
only a limited number of data points, measured at $T \ge 96$ K and $H = 
20$ kOe combine to the $M_{eff}(H)$ curve above the kink. This is why on 
the basis of the available data, no definite conclusions concerning the 
implications of this observation can be made and further studies are needed 
to clarify the situation. If this feature will be confirmed by a more 
detailed study, the kink would definitely reflect some kind of a phase 
transition.

We conclude that, contrary to the claims of Ref. \onlinecite{4}, the 
magnetization data perfectly well be explained by assuming a temperature 
independent Ginzburg-Landau parameter $\kappa$ and a conventional 
temperature dependence of the upper critical field$H_{c2}$, as shown in 
Fig. 1.


\begin{thebibliography}{71}
\bibitem{6} As is discussed in more detail in Ref. \onlinecite{5}, if the 
measurements are made at temperatures close to the superconducting critical 
temperature, the term $c_{0}(T)H$ may also account for the contribution 
to the sample magnetization arising from fluctuation effects. However, 
because this contribution is not really linear in $H$, it can only approximately 
be accounted for. At low temperatures and magnetic fields well below 
$H_{c2}$ the flactuation induced contribution to the sample 
magnetization is negligible.
\end{thebibliography}
\end{document}